\documentclass[aps,prb,reprint,groupedaddress]{revtex4-1}

\usepackage{graphics}

\begin{document}


\title{Local and long-range realizations of a  spin-reorientation surface phase transition}

\author{G. He}

\author{H. Winch}
\author{R. Belanger}
\author{P. Nguyen}
\author{D. Venus}
\email{[corresponding author] venus@physics.mcmaster.ca}
\affiliation{Department of Physics and Astronomy, McMaster University, Hamilton, Ontario, Canada}

\date{\today}

\begin{abstract}
The spin reorientation transition of an ultrathin film from perpendicular to in-plane magnetization is driven by a competition between dipole and anisotropy energies.  \textit{In situ} measurements of the magnetic susceptibility of Fe/2 ML Ni/W(110) films as a function of Fe coverage, made  as the films are deposited at constant temperature, show two clear peaks; one at the long-range and one at the local realization of the transition. In the long-range realization, the susceptibility probes the striped domain pattern that is formed in response to the balance of energetics on a mesoscopic scale.  Here the reorientation transition occurs at a non-integer layer thickness. In the local realization, the susceptibility probes the response of small islands with in-plane anisotropy in the 3rd atomic Fe layer that are  grown on the 2nd atomic Fe layer, which has perpendicular anisotropy.  It is a response to the local finite-size, metastable energetics due to discrete steps in thickness.  An excellent quantitative description of the susceptibility data is obtained when both local and long-range aspects of the spin reorientation transition are included.

\end{abstract}


\maketitle

\section{Introduction}

The spin-reorientation transition observed in ultrathin magnetic films is in reality a complex series of phenomena that illustrate the sensitive balance of exchange, dipole and anisotropy energies in a 2-dimensional magnetic system.  In simplest terms, the transition represents the canting of a uniform magnetization between orientations perpendicular and parallel to the surface of the ultrathin film\cite{[{Many examples are in: }]Bland-Heinrich,Millev1,Arnold1,Pierce1}, in response to a change in the balance of surface anisotropy and dipole energies as a function of either thickness or temperature.  In a more complete description, the perpendicularly magnetized state forms an ordered ``stripe''  pattern of domains with a pattern period that depends sensitively on the proximity of the magnetic reorientation\cite{Allenspach,Kashuba1,DeBell1}, and therefore presents an outstanding system for the controlled study of domain formation and dynamics\cite{Portman1,Kronseder1}.  Under even more detailed inspection, it has been found that the ordered domain pattern $\mathit{itself}$ undergoes a disordering transition through defect formation\cite{Abanov1,Vaterlaus1,Cannas1,Bergeard1}, and might be described by a Kosterlitz-Thouless transition.  Given this rich behaviour, it is not surprising that the spin-reorientation transition continues to provide fundamental insight into 2-dimensional magnetism after over two decades of theoretical and experimental study.

One feature that is often not sufficiently recognized is that there are two separate realizations of the spin-reorientation transition in an ultrathin film system, a long-range realization and a local realization due to finite size effects and metastablity.  Many experimental studies probe large scales where long-range dipole interactions average mesoscopically over the system.  Thus one sees reports of the continuous variation of the stripe width in the domain pattern as a function of a film thickness as the thickness changes continuously by parts of a tenth of a monolayer, even though film thickness is quantized in atomic monolayers\cite{Won1,Kronseder2,Dabrowski1,Zdyb}.  With the advent of highly spatially resolved studies using spin-polarized scanning tunnelling microscopy (STM) or low energy electron microscopy (LEEM),  it is possible to probe local metastable energetics, so that the localized effects of thickness quantization become evident.   In this case, a local realization of the transition, at a different location in parameter space, occurs\cite{Kubetzka1,Gabaly1}.

This article reports novel measurements of the magnetic susceptibility of ultrathin Fe/2 ML Ni/W(110) films as a function of coverage as the films are deposited.  Magnetic susceptibility provides complementary information to microscopy studies, because it is inherently sensitive to the dynamics of the magnetic excitations of the system, regardless of whether they are on local or long-range scales.  The present measurements clearly reveal the distinction between the local and long-range realizations of the spin-reorientation transition through the observation of two distinct peaks   in the same continuous measurement of the susceptibility on the same sample.  The two peaks are described quantitatively by models based on long-range and local processes and energetics, and yield in turn quantitative information about the long-range domain wall density and average dynamics, and the distribution of localized island sizes, respectively.  Taken together, these provide a more comprehensive, consolidated and detailed account of the series of phenomena that make up the surface spin reorientation transition.

\section{Experimental methods}
Measurements of the magnetic susceptibility were made \textit{in situ} as an ultrathin film was grown on a W(110) single crystal substrate in ultrahigh vacuum.  The sample holder\cite{Venus2} was equipped with electron beam heating for flashing to high temperature, radiative heating for temperature control, and a liquid nitrogen reservoir for cooling.  The sample could be rotated through polar and azimuthal angles, so that any in-plane crystalline axis could be aligned with an in-plane pair of magnetic field coils, and with the scattering plane of the laser beam used for the magneto-optic measurements.  For all the data reported here, the measurement axis was the W[1$\overline{1}$0] direction. A second coil attached to the holder generated a field normal to the sample surface.  The substrate cleanliness was confirmed using low energy electron diffraction and Auger electron spectroscopy (AES).

The films were formed by evaporation from a pure wire.  Electrons thermally emitted from a hot filament inside the evaporator\cite{Jones1} were accelerated by 1.75 kV and bombarded the tip of the wire.  The sublimated or evaporated atoms were collimated by two apertures and formed a beam directed at the substrate crystal.  The evaporator was supported in an adjustable tripod, so that the direction of the atomic beam could be finely adjusted and made to coincide with the region of the film probed by the laser used for magneto-optic Kerr effect (MOKE) measurements.  AES was used to iteratively adjust the evaporator direction to ensure a uniform film over a region about 9 mm$^2$ on the substrate.

The second collimating aperture in the evaporator was electrically isolated.  Because a certain fraction of the evaporate atoms striking it are ionized,  an ion current of order nA could be measured using an electrometer.  Fine adjustments of the wire position were used to keep the monitor current constant and thus ensure a constant deposition rate.  The deposition rate was calibrated by a sequence of accumulating depositions, where the film was annealed to 600K and an W Auger spectrum was measured after each step in deposition.  For Fe/W(110) and Ni/W(110), a plot of the W Auger attenuation vs. deposition time shows a clear break in slope at 1 ML that was used to calibrate the monitor current\cite{[Examples can be seen in ]Fritsch1,*Jones2}.  In the present case, the calibration constant for Fe  evaporation on W(110) was 3.00$\pm$0.15 nA min/ML.

The magnetic response of the film was determined with a MOKE apparatus\cite{Arnold2} using a linearly polarized HeNe laser.  An a.c. field of 2.0 Oe and 210 Hz was generated by either the in-plane or normal coils, depending upon the experiment.  The laser beam entered through a UHV window, scattered at 45$^o$ from the substrate normal, and exited through a second UHV window.  Compensation techniques were used to maintain linear polarization.  The beam passed through a polarizing crystal to isolate the rotated component of the light, and was detected by a photodiode.  Lock-in detection was used to isolate the signal at the frequency of the field and to measure the susceptibility directly.
%


\begin{figure}
\scalebox{.6}{\includegraphics{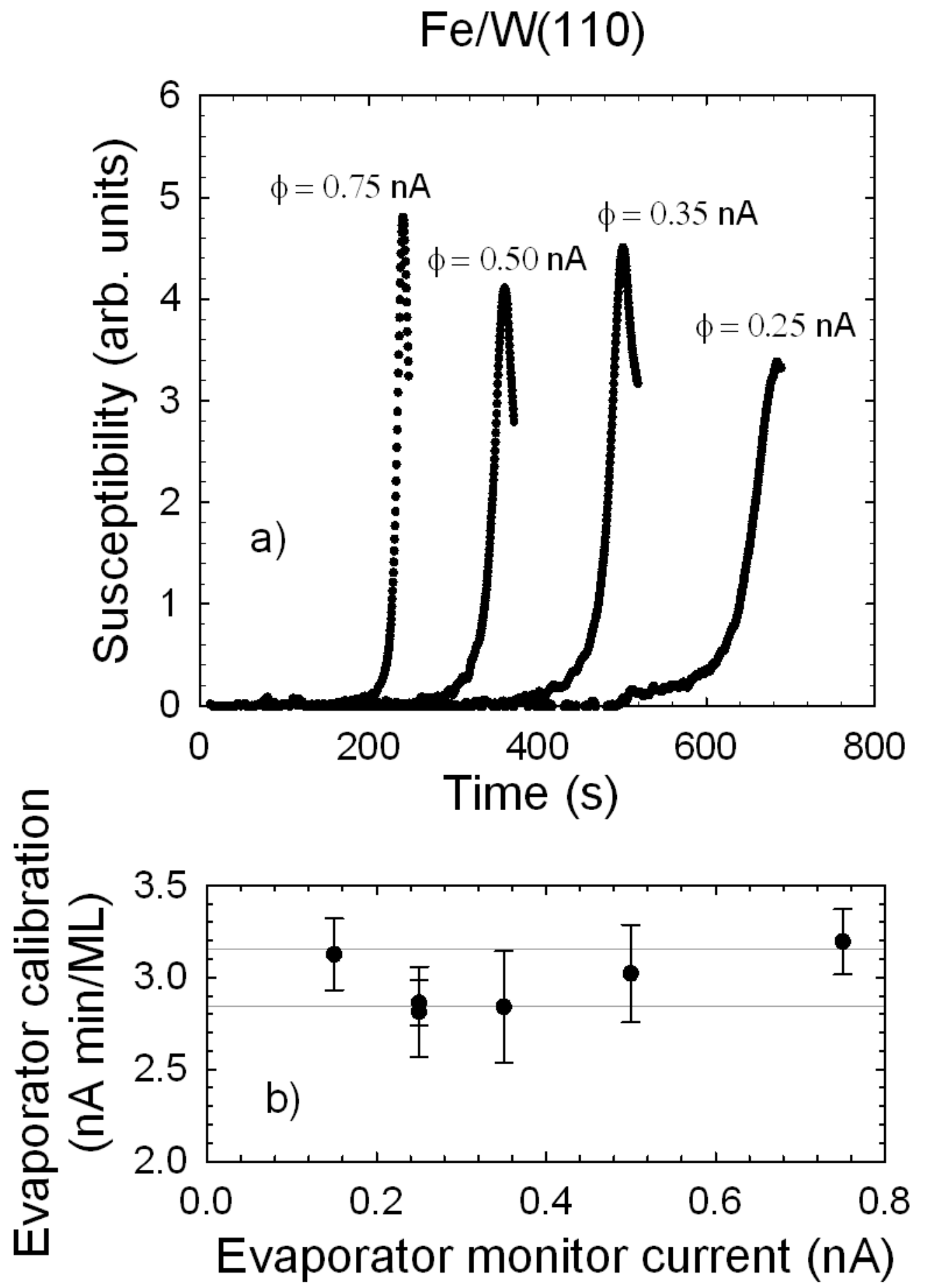}}
\caption{\label{calibration}Proof of principle and growth calibration using Fe/W(110).  a) Starting with Fe films of known thickness near 1.2 ML, the susceptibility is measured as a series of films are grown at 450 K with different evaporator monitor currents in nA.  The film undergoes a Curie transition at 2.0 ML.  b) The evaporator is calibrated using the growth time to the peak in the susceptibility at each monitor current.  The band is the error range of the independent calibration using Auger spectroscopy.   The error bar on each point reflects the uncertainty in the starting thickness of each film through the Auger thickness calibration.}
\end{figure}

The mutual alignment of the evaporator and MOKE apparatus must be rather prcise, and the use of film deposition $\theta$ as the independent variable requires growing the films very slowly.  In order to confirm the  stability of the mechanical alignment of the evaporator, and the long-term proportionality of deposition time and $\theta$, the following preliminary experiment was performed.  Fe/W(110) is known to have a Curie temperature of 450 K for films 2.0 ML thick\cite{Elmers1,Dunlavy1,Fritsch1}.  Using the evaporator calibration found via AES, a film of known  Fe thickness near 1.2 ML was grown and annealed to 600 K.  Additional deposition was then made at a fixed evaporator monitor current, $\phi$, as the susceptibility was measured using the in-plane coils.  As is shown in fig.\ref{calibration}a, the susceptibility exhibits a narrow peak at the Curie transition from paramagnetism to ferromagnetism as a function of deposition time.  Taking the deposition at the peak as 2.0 ML, and knowing the deposition time, an independent calibration of the evaporator was made at each monitor current.  These calibration factors are shown in part b) of the figure.  The six calibration points give an average calibration constant of $2.98\pm0.16$ nA min/ML, in agreement with the original AES calibration.  This shows that the evaporator is linear and reliably calibrated at low monitor currents, and that the dominant error comes from the Auger calibration method used to determine the thickness of the starting films.  With this validation of the experimental procedures, data plots in subsequent sections are presented with ML on the independent axis.

Measurements of the reorientation transition were made in an identical manner, except that the substrate was prepared with 2ML Ni/W(110).  The nickel film was annealed to 600 K after the deposition of 1 ML to cause wetting of the substrate.  In this system\cite{Johnston1}, the Ni layers create a slightly strained f.c.c. (111) surface template with atomic spacing very close to that of bulk Ni.  Subsequent pseudomorphic Fe deposition creates a system with perpendicular anisotropy for ultrathin Fe films that reorients to an in-plane magnetization along the [1$\overline{1}$0] direction of the underlying W(110) crystal.   The susceptibility was measured at room temperature using a 2.0 Oe a.c. field normal to the surface. 

\section{Long-range realization of the spin-reorientation transition}
\subsection{Analytic description}
At a first glance, it is surprising that film thickness can be treated as a continuous variable in ultrathin magnetic films, even though it is a discrete variable with large proportionate changes.   This comes about because the exchange interaction is local and (very nearly) uniform, the surface anisotropy $K_S(T)$ is (very nearly) localized to surface and interface layers, and dipole and strain energies are long-ranged and effectively average over spatial dimensions.  Under these approximations, a basic description of the spin-reorientation transition has no intrinsic scale, and the anisotropy energy $E_{an}$ depends on the surface-to-volume ratio of the film through the temperature and thickness dependence of the effective anisotropy\cite{Heinrich1}, $K_{eff}(T,\theta)$.
\begin{equation}
\label{Ean}
E_{an}=\sum_i A_i K_S(T) - \sum_i A_i d_i \Omega_D(T),
\end{equation}
where $d_i$ and $A_i$ are the thickness and surface area of the portions of the film that are $i$ atomic layers thick, and $\Omega_D(T) = \frac{1}{2} \mu_0 M^2(T)$ is the short-range dipolar demagnetization energy for a thin film geometry.
\begin{equation}
\label{Keff}
K_{eff}(T,\theta) \equiv \frac{E_{an}}{V}=\frac{K_S (T)}{b\theta}-\Omega_D(T),
\end{equation}
where the total deposition, or coverage in ML, $\theta$ is given by
\begin{equation}
\label{deftheta}
 b\theta=d=\frac{\sum_i A_i d_i}{\sum_i A_i}.
 \end{equation}
$b$ is the thickness of a single atomic layer, and $d$ is the continuous, average thickness.  In most cases where a spin-reorientation transition occurs, $K_S(T)>0$ and the system has magnetization perpendicular to the film at low thickness and/or temperature and a transition to magnetization parallel to the film surface occurs as $K_{eff}(T,\theta)$ changes sign\cite{Millev1} due to an increase in average thickness and/or thermal renormalization of the surface anisotropy with increasing temperature.  

At the next level of complexity, the long-range dipole interactions in the perpendicularly-magnetized state drive the formation of a domain pattern with equilibrium stripe density\cite{Kashuba1,Abanov1}
\begin{equation}
\label{n}
n^{eq}(T,\theta)=\frac{4}{\pi \ell(T,\theta)} \exp\Bigl(-\frac{E_W(T,\theta)}{4\Omega_D(T) b\theta}-1\Bigr),
\end{equation}
where $\ell=\pi [\Gamma/K_{eff}(T,\theta)]^{\frac{1}{2}}$ and $E_{W}=4[\Gamma K_{eff}(T,\theta)]^{\frac{1}{2}}$ are the domain wall width and energy/area.  $\Gamma$ is the exchange stiffness and the use of the effective anisotropy and average thickness is justified by the long-range nature of the dipole interactions.  In the presence of the stripe domain phase, applying a magnetic field along the surface normal causes the domains to grow (shrink) if the domain magnetization is parallel (antiparallel) to the applied field.  The equilibrium magnetic susceptibility $\chi^{eq}_{\bot}(T,\theta)$, measures the moment induced in this way\cite{Libdeh1}.
\begin{equation}
\label{chieq}
\chi^{eq}_{\bot}(T,\theta)=\frac{2}{\pi^2d}\frac{1}{n^{eq}(T,\theta)}=\frac{\ell}{2\pi b\theta}\exp\Bigl(\frac{E_W(T,\theta)}{4\Omega_D(T) b\theta}+1\Bigr).
\end{equation}

The domain walls in the domain pattern are subject to pinning by inhomogeneities in the film, so that the domains respond to the applied field with a relaxation time $\tau$. The relaxation time depends upon the  depth, or activation energy $E_a$, of the local pinning potential and on an intrinsic ``attempt'' time $\tau_0$ for the trapped domain wall segment to escape.
\begin{equation}
\label{tau}
\tau=\tau_0 \exp\Bigl(\frac{E_a}{kT}\Bigr).
\end{equation}
The  pinning sites represent a distribution of activation energies. Bruno \textit{et al.}\cite{Bruno1} have considered the question of estimating the mean activation energy due to variations in film thickness.  They use a spatial averaging technique to reduce the two-dimensional pinning problem to that of a rigid domain wall moving in an one-dimensional effective potential.  Due to the spatial averaging, the resulting mean activation energy depends upon the continuous, average film thickness, $d$.
\begin{equation}
\label{Eadef}
E_a(T,d)=\frac{\zeta d}{E_W(T,d)} \Bigl(\frac{\partial}{\partial d} E_W(T,d) \Delta d\Bigr)^2,
\end{equation}
where $\zeta$ is the mean spacing of pinning sites.  Because the pinning is most effective at lower thickness where $K_{eff}\not\approx0$, $\frac{\Omega_D(T) d}{K_S}$ can be treated as a small parameter in the regime where pinning is effective.  Choosing $\Delta d$ to be a layer thickness $b$,
\begin{equation}
\label{Ea}
E_a(T,\theta)\approx \zeta \sqrt{\Gamma K_S(T)b}\: \theta^{-\frac{3}{2}}\Bigl(1 + \frac{3}{2}\frac{\Omega_D(T) b\theta}{K_S(T)}\Bigr).
\end{equation}

The dynamics of the domain walls in an a.c. applied field with a low frequency $\omega$ can be taken treated in the relaxation approximation. This gives
\begin{equation}
\label{chi}
\chi_{\bot}(T,\theta)=\frac{1-i\omega\tau(T,\theta)}{1+\omega^2\tau^2(T,\theta)} \; \chi^{eq}_{\bot}(T,\theta)
\end{equation}
The above model has been previously used to describe susceptibility measurements of perpendicularly magnetized ultrathin films approaching the reorientation as a function of \textit{temperature}\cite{Venus1}.  This has provided complementary information to microscopy studies, by describing the dynamics of domain wall formation and pattern defects in the domain pattern\cite{Libdeh1,Libdeh2}.
 In the next sections, the model is applied to novel measurements of the susceptibility as a function of $\theta$.

\subsection{Long-range reorientation as a function of $\theta$}
\begin{figure}
\scalebox{.8}{\includegraphics{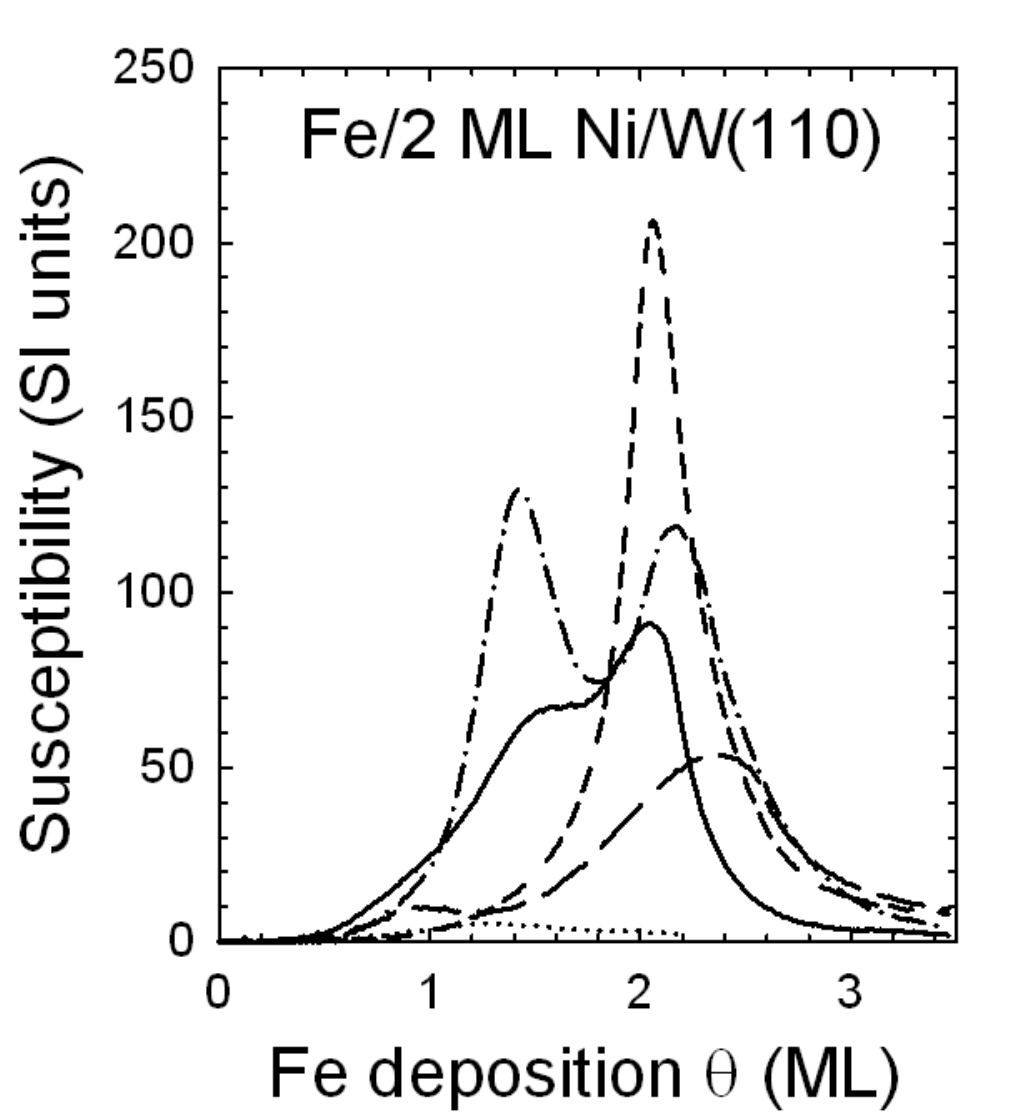}}
\caption{\label{overview}A representative selection of susceptibility measurements of Fe/2 ML Ni/W(110) measured as a function of coverage as the films were being grown at room temperature.}
\end{figure}

A collection of measurements of $\chi_{\bot}(\theta)$ for Fe/2 ML Ni/W(110) films is shown in fig. \ref{overview}.  One sees immediately that there are usually two peaks in the susceptibility, even though previous studies\cite{Arnold3,Dunlavy2} of films of fixed thickness as a function of temperature implied that there would be only one peak preceding the reorientation that occurs past $\theta$=2 ML.  The new peak at lower coverage is particularly sensitive to the growth conditions.  This peak sometimes appears as a well-resolved peak (dash-dot curve), sometimes as a shoulder (solid curve), sometimes as very small peak (long dash curve), and is occasionally not clearly present (short dash curve).  It was possible to measure a susceptibility curve where only the first peak was present (dotted curve) by growing on an unannealled Ni film to create a particularly rough Fe film, but then the first peak was very small.  This leads to the hypothesis that the peak at higher deposition corresponds to that seen in previous studies as a function of temperature, and represents the response of the stripe domain pattern which precedes the long-range reorientation.  It is absent for the very rough film because the domain walls in the stripe domain pattern remain pinned at all coverages.

\begin{figure}
\scalebox{.6}{\includegraphics{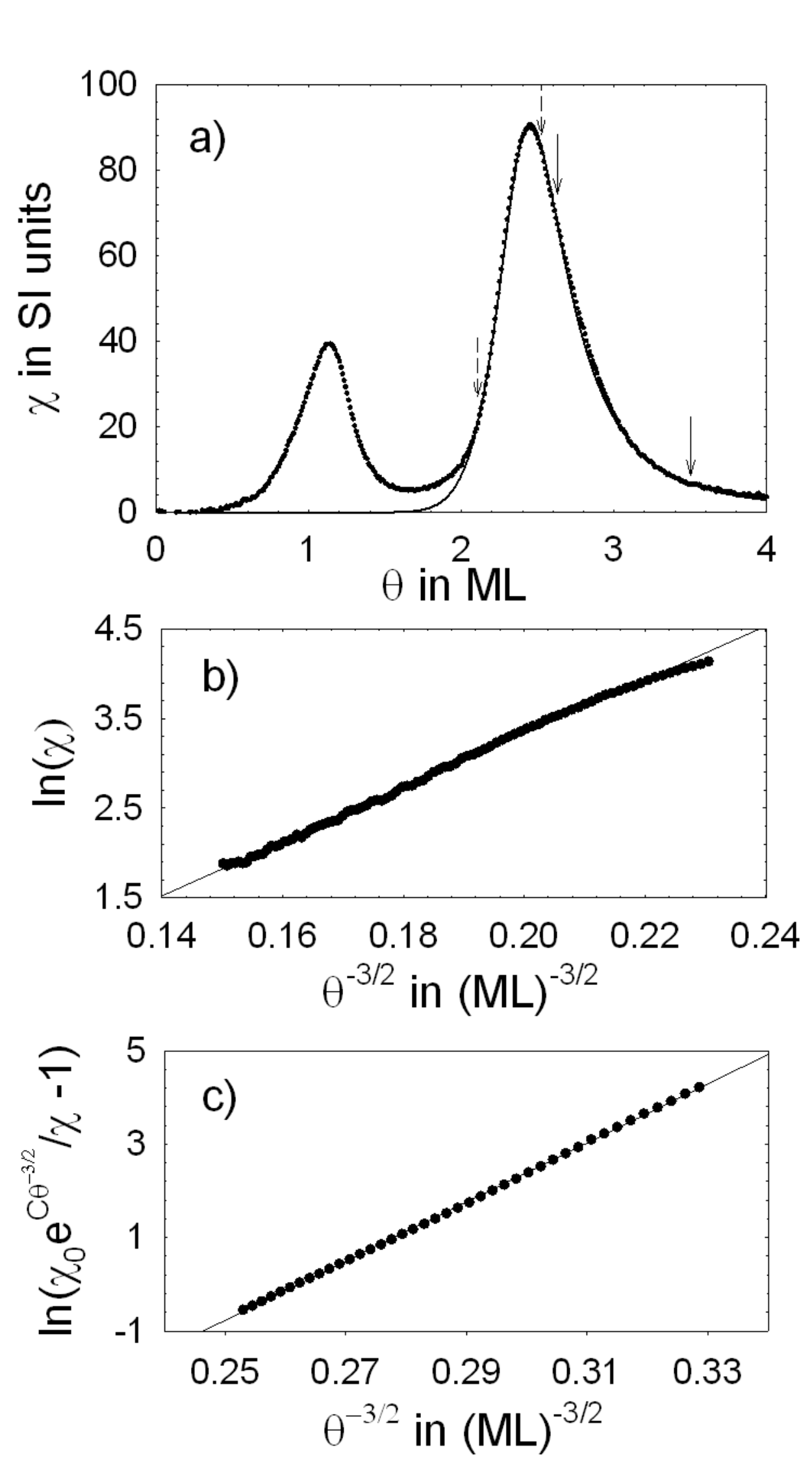}}
\caption{\label{fits}Representative analysis of the second peak in the susceptibility, using the long-range domain model. a) Susceptibility of a sample with well separated peaks.  The solid line, mostly obscured by the data symbols, is obtained by fitting to the model in section IIA.  b) The section of the curve in part a) between the solid arrows is fitted to $\theta^{-\frac{3}{2}}$, in accordance with eq.(\ref{chieq}). c) The section of the curve in part a) between the dashed arrows is fitted to $\theta^{-\frac{3}{2}}$, in accordance with eq.(\ref{tau}), (\ref{Ea}) and (\ref{chi}). The constants $\chi_0$ and $C$ are determined from the fit in part b).}
\end{figure}

In order to test this hypothesis, three measurements where the two peaks are well separated were selected for quantitative analysis.  A sample analysis of the peak at larger $\theta$, using the model for a long-range reorientation transition, is shown in fig. \ref{fits}.  The fitted curve is mostly obscured by the data points (which have been binned in increments of 0.01 ML).   The pinning energies are reduced at high coverage, and the dynamical prefactor in eq.(\ref{chi}) can be neglected.  According to eq.(\ref{chieq}), the logarithm of this portion of the curve should vary as $\theta^{-\frac{3}{2}}$ to first order.  This is demonstrated in part b) for the data between the solid arrows.  A linear least-squares fit provides the constants $\ln\chi_0$ and $C$.  On the low coverage side of the peak, the dynamical factor grows as a result of domain wall pinning at lower coverage.   This is described by eq.(\ref{tau}), (\ref{Ea}) and (\ref{chi}).  Using the constants from the fit in part b), the scaling of the average activation energy as $\theta^{-\frac{3}{2}}$ is illustrated in fig.(\ref{fits}c) for the data between the two dashed arrows.  The four constants determined from the two least-squares fits give an excellent description of the peak.

\begin{figure}
\begin{center}
\scalebox{.6}{\includegraphics{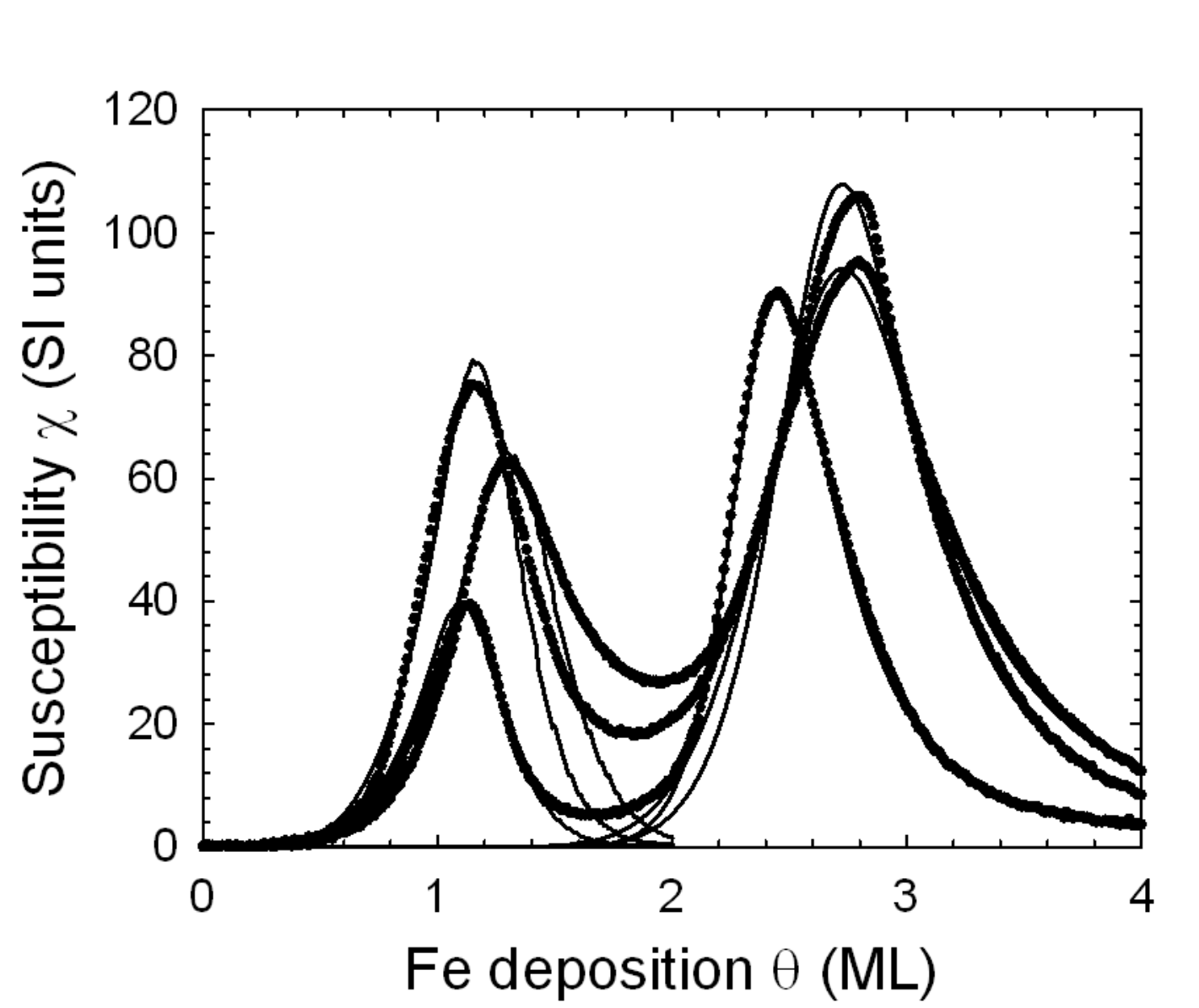}}
\caption{\label{3 sets}Selected susceptibility measurements where the two peaks are well separated.  The curves fit to the peaks at higher deposition are based on the model of domain motion in the striped domain pattern that precedes the long-range reorientation transition.  The curves fit to the peaks at lower deposition are discussed in section III.}
\end{center}
\end{figure}

Figure \ref{3 sets} shows that the peak at high deposition in all three of the data sets with well-separated peaks are described very well by the model of the stripe domain pattern that accompanies the long-range realization of the reorientation transition.  Microscopy studies showing that the domain density changes exponentially with thickness\cite{Won1,Dabrowski1,Zdyb} are confirmed, with the further finding that the argument of the exponential has a leading term that varies as $\theta^{-3/2}$, as predicted.  The data sets with a larger signal in the ``valley'' between the two peaks likely represent situations where the distribution of pinning sites is broader.   For this reason, the model based on the mean pinning energy does not fit the tail at low deposition quite as well.  Nevertheless, it is clear that the one-dimensional effective theory of pinning by thickness variations is very successful, even in the ultrathin film limit.  A detailed analysis of the fitting constants, and what they reveal about the mean separation of the pinning sites, $\zeta$, and the magnetic properties of the films, is left to a later publication that considers a larger collection of measurements across a range of temperatures.  The present article turns instead to the new peak at lower deposition.   


\section{Local realization of the reorientation transition}
\subsection{Analytic description}
A successful description of the first peak in the susceptibility must be consistent with a number of observations.  First, it cannot involve the motion of domain walls, since these are pinned at all coverages less than the peak at larger $\theta$.  The first peak must involve a different low energy excitation, or ``soft'' mode, that will respond to a small oscillating perpendicular field.  Second, the peak is relatively narrow, occupying a small range of $\theta$ near, but significantly above, 1 ML.  Third, although the peak is sensitive to the growth conditions, it persists even for very rough films.  All these points suggest the magnetic response of islands.  There are two possibilities:  the peak might be the response of small 1st monolayer Fe islands with perpendicular anisotropy surrounded by the in-plane Ni substrate film;  or the peak might represent the response of Fe islands with in-plane anisotropy in the 3rd atomic layer, that lie within much larger regions of 2 atomic layer Fe thickness with perpendicular anisotropy.  If the first scenario were correct, one would expect the response to begin at much lower coverage than it does, and to reach a maximum well below a coverage of 1 ML.  For this reason, the following analytic description is presented with reference to the islands in the 3rd Fe layer.  However, the model can be easily adapted to the first scenario, and this question will be revisited.

A number of authors have considered an analytic description of the formation of a static domain wall at the interface between two regions with different axes of magnetic anisotropy\cite{Elmers2,Elmers3,Weber1}.  Although this description is tractable in one dimension (infinite, uniform stripes), it is much more difficult for two dimensional islands.  A simplified approach that captures the essence of the situation is to assume that there is a critical island radius, $r_c$.  Islands in the 3rd layer that have radius $r<r_c$ have a partially-developed domain wall at the interface with the 2 atomic layer region within which they lie.  This spin geometry is a compromise between the competing exchange energy with the surrounding 2 atomic layer region and the in-plane anisotropy energy within the island, and represents a system in transition that is sensitive, or ``soft'', to a small applied perpendicular field.  Islands with $r>r_c$ have a fully formed domain wall at their boundary, but the domain wall is pinned and cannot move.  In the interior of the islands the in-plane anisotropy and exchange energies are mutually reinforcing, rather than competing, so that there is no significant response to a small perpendicular field.   

This simple model is equivalent to determining the net magnetic moment from  islands in the 3rd atomic layer below a certain size, and how it responds to an applied perpendicular field.  Islands in the 3rd layer grow on larger islands in the 2nd layer, and have in turn islands in the 4th layer growing on top of them.  Let $s_3$ be the number of atoms in a particular island in the 3rd layer of a film, and $N(s_3)$ be the site density of islands in the 3rd layer containing $s_3$ atoms.  The critical island size is when $s_3=s_3^*$.  Then the total site density of islands in the 3rd layer is $N_3=\sum_{s_3} N(s_3).$  The magnetization that results if the moments of all portions of the film where the surface atoms are in the 3rd layer are aligned is
\begin{equation}
M_{max}=(\frac{1}{a_0^2 3b})\Bigl(\sum_{s_3 <s_3^*}N(s_3) \sum_{s_4<s_3}\frac{N(s_4)}{N_4}\Bigr)(3\mu(s_3-s_4)).
\label{Mmax}
\end{equation}
The first term is the (site/volume).  The second is the number of (islands/site), which has been corrected for the fraction of each island in the 3rd layer that is covered by a 4th layer island.  The final term is the (moment/island).  $a_0$ is the in-plane lattice constant, and $\mu$ is the moment/atom.  The degree to which the spins in the partial domain wall in each island can be aligned by an external field $H$ at temperature $T$ is approximately given by multiplying each term under the double sum over $s_3$ and $s_4$ in eq.(\ref{Mmax}) by the low-field limit of the Langevin function\cite{Chikazumi}:
\begin{equation}
 \frac{1}{3}\Bigl(\frac{3\mu_0\mu}{kT}(s_3-s_4)H\Bigr).
\label{detailed}
\end{equation}
Then the susceptibility, in the limit of the applied field going to zero, is
\begin{equation}
\chi=\frac{2a_0^2b}{kT}\Omega_D \sum_{s_3<s^*}N(s_3) \sum_{s_4<s_3}\frac{N(s_4)}{N_4}(s_3-s_4)^2,
\label{chiraw}
\end{equation}
where it has been recognized that $\frac{1}{2}\mu_0(\frac{\mu}{a_0^2b})^2=\Omega_D$

Amar and Family\cite{Amar2} have presented a model of the island size distribution as a monolayer film grows.  It uses the general scaling condition
\begin{equation}
N(s_i)=\frac{N_i^2}{\theta_i} f(u_i),
\label{islands}
\end{equation}
where $\theta_i$ is the coverage in the $i^{th}$ layer, such that $\theta=\sum_i \theta_i$.  $f(u_i)$ is a scaling function in the dimensionless variable $u_i=\frac{s_i}{\sigma_i}$, where $\sigma_i$ is the average number of atoms contained in each island in the $i^{th}$ layer.  An expression for the scaling function is given in ref. \onlinecite{Amar2}.  It has normalization
\begin{eqnarray}
\int_0^\infty f(u) du = \int_0^\infty u f(u) du = 1 \\
  \int_0^\infty u^2 f(u) du = 1.09 \equiv c_2.
\label{norm}
\end{eqnarray}
The function $f(u)$ is such that these integrals are insensitive to the value of the upper limit so long as it is $\geq2$. Substituting these definitions into eq.(\ref{chiraw}), and converting the sums to integrals over the continuous variables $u_i$, gives
\begin{equation}
\chi=\frac{2a_0^2b}{kT}\Omega_D \int_0^{u_3^*}du_3 N_3 f(u_3) \int_0^{u_4^{m}} du_4 f(u_4)(\frac{u_3\theta_3}{N_3}-\frac{u_4\theta_4}{N_4})^2,
\label{integrals}
\end{equation}
where $u_4^m$ represents the largest island size in the 4th layer, and $u_3^*$ represents the critical island size in the 3rd layer.  

It is now assumed that the island growth in the 3rd and 4th layers is beyond the nucleation stage, but still far from layer percolation.  Newly deposited atoms are much more likely to aggregate on an existing island than to nucleate a new island.  In this aggregation stage\cite{Amar1}, the number of islands remains constant.  Thus
\begin{equation}
\sigma_i=\frac{\theta_i}{N_i}.
\end{equation}
In addition, since both the 3rd and 4th layers are described by monolayer Fe islands growing on existing Fe islands, the island density after nucleation has ended will be the same, $N_i=N$.  Under these conditions, the upper limits of the integrals can be written as
\begin{eqnarray}
u_3^* & = & \frac{s_3^*}{\sigma_3}=\frac{s_3^*}{\theta_3}N \equiv \frac{\theta_3^*}{\theta_3} \\
\label{upper3}
u_4^{m} & = & \frac{s_4^{m}}{\sigma_4}=\frac{s_3}{\theta_4}N=\frac{s_3}{\theta_4}\frac{\theta_3}{\sigma_3} = u_3 \frac{\theta_3}{\theta_4},
\label{upper4}
\end{eqnarray}
and the susceptibility becomes
\begin{eqnarray}
\label{full chi}
\chi & = & B\int_0^{\frac{\theta_3^*}{\theta_3}} du_3 f(u_3) \int_0^{u_3\frac{\theta_3}{\theta_4}} du_4 f(u_4) (u_3\theta_3 - u_4\theta_4)^2 \\
B & = & \frac{2a_0^2b}{NkT}\Omega_D.
\label{B}
\end{eqnarray}
When this expression is used to describe small 3rd layer islands with radius $r<r_c$, the 4th layer islands growing on top of them will be very small. Then $u_4^{m}\sim \frac{\theta_3}{\theta_4} \gg2$, and the upper limit in eq.(\ref{full chi}) can be replaced by infinity.  The final expression for the susceptibility is
\begin{equation}
\chi(\theta)=B\int_0^{\frac{\theta_3^*}{\theta_3}} du_3 f(u_3) (u_3^2\theta_3^2-2u_3\theta_3\theta_4+c_2\theta_4^2).
\label{chifinal}
\end{equation}

To evaluate this requires a model that relates the total coverage $\theta$ to the layer coverages $\theta_i$.  The most straightforward approach, which is valid at intermediate temperatures, is to assume that newly deposited atoms have sufficient thermal energy to aggregate, but not enough to move from the layer upon which they are deposited to another layer.  This condition is described by the differential equation
\begin{equation}
d\theta_i=(\theta_{i-1}-\theta_i) d\theta,
\label{diffeq}
\end{equation}
which merely states that the $i^{th}$ layer must grow on uncovered regions of the $(i-1)^{th}$ layer.  It has the solution
\begin{equation}
\theta_i = 1-e^{-\theta}\sum_{n=0}^{i-1}\frac{\theta^n}{n!}.
\label{solution}
\end{equation}
This can be substituted into each term in eq.(\ref{chifinal}) for numerical integration.  In the limit of low coverages in the 3rd layer, $\frac{\theta_3^*}{\theta_3}>2$ and the upper limit of the integral can be set to infinity. Then
\begin{equation}
\chi(\theta)=B(c_2\theta_3^2-2\theta_3\theta_4+c_2\theta_4^2)\approx B(\theta_3-\theta_4)^2=B\frac{\theta^6}{36}e^{-2\theta}. \\
\label{theta6}
\end{equation}
This expression should describe the leading edge of the first susceptibility peak.  Recall at this point that it is also conceivable that this peak in the susceptibility represents the response of small islands with perpendicular anisotropy in the first Fe monolayer, that are surrounded by the in-plane Ni substrate.  In this case, the leading edge of the peak would vary as
\begin{equation}
\chi(\theta)\sim (\theta_1-\theta_2)^2 = \theta^2 e^{-2\theta}.
\label{theta2}
\end{equation}

\subsection{Local reorientation of islands as a function of $\theta$}
\begin{figure}
\scalebox{.7}{\includegraphics{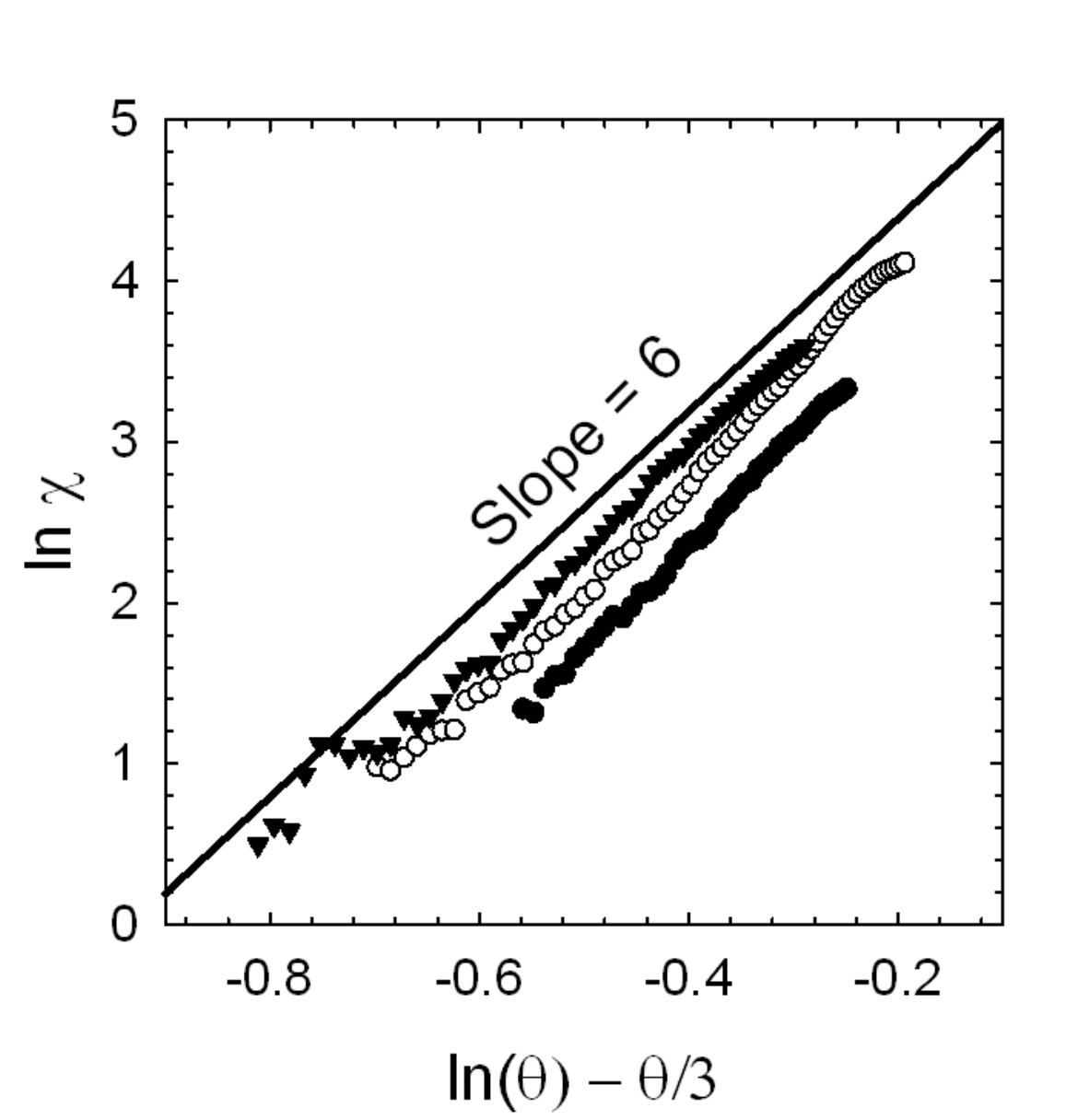}}
\caption{\label{lntheta}Scaling of the leading edge of the first peak in the data sets in fig.(\ref{3 sets}).  According to eq.(\ref{theta6}), the slope should equal 6 if the susceptibility is the response of small islands in the 3rd layer.}
\end{figure}

The model of local reorientation is tested by applying it to the first peak  of the data sets in fig.(\ref{3 sets}).   In accordance with eq.(\ref{theta6}), the leading edge of the peak with $\theta \leq 1 ML$  is plotted in fig.(\ref{lntheta}), with $\ln\chi$ as a function of $\ln\theta - \frac{\theta}{3}$.
The fact that the slope on all three of the plots is very nearly 6 is strong evidence that the magnetic response is from small islands of 3 layer thickness.  It may be that the alternate hypothesis of a peak due to islands in the first Fe layer will be observable as yet another peak at lower coverage under different experimental conditions.  Once the intercept on the plot is used to determine the prefactor B in eq.(\ref{B}), the only remaining free parameter is $\theta_3^*$.  This can be determined by calculating $\chi(\theta)$ in eq.(\ref{chifinal}) for a range of values of $\theta_3^*$ and finding the value that gives the minimum least-squares deviation from the data in the region of the top of the first peak.  Curves fitted in this way are included in fig.(\ref{3 sets}).  The curves give an excellent representation of the data, and are for the most part obscured by the data points.  Again, in the ``valley'' between the two peaks the curves depart significantly from the data.  This is likely a result of using a uniform cut-off to the integral, rather than a more nuanced approach that might take into account variables such as island shape, and of course due to overlap with the tail of the second peak.
%
\begin{table}
\caption{\label{parameters}The model for the local reorientation of islands in the 3rd layer requires the fitting of the two free parameters, $B$ and $\theta_3^*$.  From these, the physical parameters for the island density $N$, the critical size $s_3^*$ and diameter $r_c$ of the islands where there the local transition is complete, can be derived.}
\begin{ruledtabular}
\begin{tabular}{|l|l|l|l|}
Data set          &         1     &     2      &     3         \\ \hline
$B$ (SI units)     & 7,670     &  6,190 &  12,230  \\ \hline
$\theta_3^*$ (atoms/site)  &  0.145  & 0.210  &  0.165 \\ \hline \hline
$N/a_0^2$ (islands/$\mu$m$^2)$                    & 23  &  29 &   15 \\ \hline
$a_0^2 s_3^*$ ($\mu$m$^2$/island)     & 6.2$\times10^{-3}$     & 7.2$\times10^{-3}$  & 11.1$\times10^{-3}$ \\ \hline
$r_c$ (nm)                      & 45    &  48   &  59 \\ 
\end{tabular}
\end{ruledtabular}
\end{table}

The fitted values of the parameters $B$ and $\theta_3^*$ are collected in Table \ref{parameters}.  The value of $\theta_3^*$ is well within the aggregation regime of the island growth model\cite{Amar1}, as was assumed.  It can also be used to check the assumption that was made in moving from eq.(\ref{full chi}) to eq.(\ref{chifinal}).  By rearranging eq.(\ref{upper4}), the condition $u_4^{m}>2$ becomes $u_3>2\frac{\theta_4}{\theta_3}$.  The most stringent test is at the largest value of $\theta$, when $\theta_3=\theta_3^*\approx0.2$.  In this case, eq.(\ref{solution}) gives $\frac{\theta_4}{\theta_3}\approx1/3$, so that $u_3>0.67$.  This condition omits a portion of the integral in eq.(\ref{full chi}) at small values of $u_3$.  Recalculating the normalization conditions in eq.(\ref{norm}) from this lower limit (rather than 0) shows that a maximum error of 10\% is created. This error occurs at the point where the fitted curve is deviating from the data because of the ``valley'' between the two peaks in any case.  It is concluded that the assumption does not significantly effect the analysis. 

The fitted parameters can be related to physical parameters describing the distribution of islands in the 3rd layer.  The island density $N/a_0^2$ can be found from eq.(\ref{B}), using the values $\Omega_D=1.9\times10^6$ J/m$^3$ for b.c.c. iron\cite{Chikazumi}, and $b=0.20$ nm for the Fe lattice formed on the 2 ML Ni f.c.c. substrate\cite{Johnston1}.  The critical island area $a_0^2s_3^*$ can then be found using eq.(\ref{upper3}), and the radius of  the (circular) critical islands, $r_c$, calculated.  An important point of internal consistency is to compare the calculated critical island radius to the assumptions of the model.  According to the model, the magnetic response of the 3rd layer islands to a small perpendicular field disappears once a complete, pinned domain wall is formed to transition from perpendicular moments at the edge of the island to in-plane moments at the centre.  Thus the critical island radius should be about the length of a $90^o$ domain wall.   The entries in Table \ref{parameters} give a average critical island radius of $\approx$ 50 nm.  This is very reasonable for the width of a domain wall.

\section{Conclusions}
Novel \textit{in situ} measurements of the magnetic susceptibility of Fe/2 ML Ni/W(110) films, made as a function of deposition as the films were grown, have provided a more comprehensive view of the collection of related phenomena that are referred to as the surface spin-reorientation transition.  The measurements reveal two clear peaks in the susceptibility associated with the transition, rather than the single peak that might have been inferred from microcopy studies or studies as a function of temperature.  The peak at lower coverage is sensitive to details of the film growth, and its precise form is not routinely reproducible from film to film.  However, in instances where the two peaks are well-separated they can be separately modelled and analysed.

The peak at larger deposition ($\theta \geq$ 2 Fe ML) results from the long-range realization of the reorientation transition, where the energetics are governed by mesoscopic averaging over the film morphology and lead to long-range domain structures.  This signature peak is well-described by an analysis in terms of the continuous average film thickness, or total coverage, $\theta$.  The peak is a response to the motion of domain walls in the stripe domain pattern.  The domain density is found to be an exponential function of average film thickness, in agreement with previous microscopy studies.  It is further shown that the functional form of the argument of the exponential has a leading term that varies as $\theta^{-3/2}$, in agreement with theory.  The domain walls are pinned with an activation energy that also varies as $\theta^{-3/2}$, but through a different mechanism.  This variation results from the mean of the distribution of the averaged effective pinning potential due to statistical variations in film thickness.  The model gives an excellent description of the magnetic susceptibility with only 4 independent parameters.  A more extensive investigation of the long-range thickness-dependent transition, including the extraction of relevant properties of the magnetic system from the fitted parameters, will be the subject of a future article.

The peak at lower deposition ($\theta \leq $1.5 Fe ML) results from a local realization of the reorientation transition, where the effects of finite size and metastability are important.  As a result, the reorientation must be described in terms of local, discrete layer thickness and coverages $\theta_i$ rather than the total coverage. The discrete step in thickness between the 2nd and 3rd layers creates a boundary between regions with perpendicular and in-plane anisotropy.  The partial domain wall pinned at this boundary on islands smaller than a critical radius is susceptible to a small perpendicular field.  The model of the response of the islands in a local realization of the transition has only two free parameters.  First, the leading edge of the susceptibility peak scales as $\chi(\theta)\sim\theta^6\exp({-2\theta})$, in agreement with the aggregation of small 3rd layer islands on the 2nd layer.  Second, the response disappears once the 3rd layer islands have a radius larger than $\approx$ 50 nm, which is a consistent with the formation of a complete pinned domain wall at the island boundary.

These results make it clear that both the local and long-range realizations of the reorientation transition are needed to understand the magnetic response of the system.  The two realizations are distinct in that they involve different magnetic excitations, different descriptions of the film morphology, and occur at different locations in parameter space.  They are tied together within the group of related phenomena that result from the re-balancing of dipole energy and anisotropy energy as the temperature and/or average thickness of an ultrathin film is varied.



\bibliography{reorientation}

\end{document}